\documentclass[aps,prb,twocolumn,10pt,superscriptaddress,showpacs]{revtex4-2}

\usepackage{hyperref}
\usepackage{epsfig}
\usepackage{graphicx}
\usepackage{subfigure}
\usepackage{latexsym}
\usepackage{color}
\usepackage{fullpage}
\usepackage{dcolumn}
\usepackage{bm}
\usepackage[normalem]{ulem}
\usepackage{amsmath}
\usepackage[paperwidth=210mm,paperheight=297mm,centering,hmargin=2cm,vmargin=2.3cm]{geometry}

\usepackage{booktabs}
\usepackage{siunitx}

\begin{document}

\title{Pressure-induced electronic and structural evolution of EuIrGe$_3$}

\author{N. S. Dhami}
\email{naveen-singh.dhami@synchrotron-soleil.fr}
\affiliation{Institute of Physics, Bijeni\v{c}ka cesta 46, 10000, Zagreb, Croatia}
\affiliation{Synchrotron SOLEIL, L’Orme des Merisiers, Saint Aubin BP 48, 91192 Gif-sur-Yvette, France}

\author{V. Balédent}
\affiliation{Université Paris-Saclay, CNRS, Laboratoire de Physique des Solides, UMR-8502, 91405, Orsay, France}
\affiliation{Institut universitaire de France (IUF), Paris, France}

\author{ I. Batistić}
\affiliation{Department of Physics, Faculty of Science, University of Zagreb, Bijeni\v{c}ka 32, 10000 Zagreb, Croatia}

\author{C. M. N. Kumar}
\affiliation{The Henryk Niewodniczański, Institute of Nuclear Physics, Polish Academy of Sciences, ul. Radzikowskiego 152, 31-342 Kraków, Poland}

\author{D. Kaczorowski}
\affiliation{Institute of Low Temperature and Structure Research, Polish Academy of Sciences, Okólna 2, 50-422 Wrocław, Poland}

\author{L. Nataf}
\affiliation{Synchrotron SOLEIL, L’Orme des Merisiers, Saint Aubin BP 48, 91192 Gif-sur-Yvette, France}

\author{J. M. Ablett}
\affiliation{Synchrotron SOLEIL, L’Orme des Merisiers, Saint Aubin BP 48, 91192 Gif-sur-Yvette, France}

\author{J.-P. Rueff}
\affiliation{Synchrotron SOLEIL, L’Orme des Merisiers, Saint Aubin BP 48, 91192 Gif-sur-Yvette, France}
\affiliation{Laboratoire de Chimie Physique-Matière et Rayonnement, Sorbonne Université, CNRS, 75005 Paris, France}

\author{J. P. Itié}
\affiliation{Synchrotron SOLEIL, L’Orme des Merisiers, Saint Aubin BP 48, 91192 Gif-sur-Yvette, France}

\author{P. Fertey}
\affiliation{Synchrotron SOLEIL, L’Orme des Merisiers, Saint Aubin BP 48, 91192 Gif-sur-Yvette, France}

\author{S. R. Shieh}
\affiliation{Department of Earth Sciences, Department of Physics and Astronomy, University of Western Ontario, London, Ontario N6A-5B7, Canada}

\author{P. Popčević}
\affiliation{Institute of Physics, Bijeni\v{c}ka cesta 46, 10000, Zagreb, Croatia}

\author{Y. Utsumi Boucher}
\email{yutsumi@ifs.hr}
\affiliation{Institute of Physics, Bijeni\v{c}ka cesta 46, 10000, Zagreb, Croatia}

\begin{abstract}
We investigated the pressure-induced evolution of the electronic and crystal structure of the noncentrosymmetric BaNiSn$_3$-type antiferromagnet EuIrGe$_3$ using x-ray absorption spectroscopy, synchrotron x-ray diffraction complemented by density functional theory calculations, and electrical resistivity measurements. The Eu $L_3$-edge spectra reveal a continuous increase in the mean Eu valence under compression, accompanied by modifications of the Ge and Ir electronic states. High-pressure x-ray diffraction shows anisotropic lattice compression in the tetragonal ($I4mm$) phase and provides evidence for a structural phase transition above 38 GPa. The experimentally determined lattice and equation-of-state parameters are in good agreement with the DFT calculations. Electrical resistivity measurements reveal a monotonic increase in the antiferromagnetic ordering temperatures up to 18 GPa, indicating that the antiferromagnetic ground state remains robust despite the increasing contribution of the nonmagnetic Eu$^{3+}$ configuration to the intermediate valence state. These results demonstrate that EuIrGe$_3$ exhibits a pressure response distinct from EuCoGe$_3$ and EuRhGe$_3$, highlighting the important role of the transition metal $d$-electron states in the pressure-induced electronic and structural evolution of the Eu$T$Ge$_3$ family.

\end{abstract}

\date{\today}

\maketitle

\section{Introduction}

Intermetallic Eu-compounds have attracted considerable attention because of their rich magnetic and electronic properties. Many ternary Eu-compounds crystallize in BaAl$_4$-derived crystal structures, most notably the centrosymmetric ThCr$_2$Si$_2$-type and the noncentrosymmetric BaNiSn$_3$-type structures. Both structural families commonly exhibit antiferromagnetic ordering of Eu$^{2+}$ (4$f^7$) moments mediated by the Ruderman-Kittel-Kasuya-Yosida (RKKY) interaction. However, several members exhibit temperature-, pressure-, or chemical- substitution-induced valence instabilities, resulting in intermediate-valence states or valence transitions toward the nonmagnetic Eu$^{3+}$ (4$f^6$) state \cite{Onuki_PM_2017, Onuki2020}. 
 
Interestingly, despite their common Eu$^{2+}$-based magnetism, the pressure response differs markedly between these two structural families. ThCr$_2$Si$_2$-type Eu-compounds, such as EuCo$_2$Ge$_2$  ($T_{\rm N}\sim$23 K) and EuNi$_2$Ge$_2$  ($T_{\rm N}\sim$30 K) \cite{Felner1978}, exhibit pressure-induced valence transitions at relatively low pressure, accompanied by a rapid suppression of antiferromagnetic order and the development of intermediate valence or nearly Eu$^{3+}$ states \cite{Hess1997, Dionicio2006, Muthu2016}. In contrast, BaNiSn$_3$-type Eu-compounds show a more gradual pressure dependence of the Eu valence, with antiferromagnetic ordering persisting above 8 GPa \cite{Kakihana2017, Muthu2019,Utsumi_ES_2021,chen2023evidence, dhami2023pressure}.

Even within the BaNiSn$_3$-type Eu$T$Ge$_3$ family, the pressure response varies significantly depending on the transition metal element $T$, indicating an important role of the transition metal $d$-electron states in governing the electronic and structural evolution under pressure. Our previous pressure-dependent x-ray absorption spectroscopy (XAS) and x-ray diffraction (XRD) studies revealed that EuCoGe$_3$ and EuRhGe$_3$ both exhibit anisotropic lattice compression without evidence of a structural phase transition \cite{dhami2023pressure, dhami2024synchrotron}. However, their Eu valence evolves differently under pressure, EuCoGe$_3$ remains close to Eu$^{2+}$ up to 50 GPa, whereas EuRhGe$_3$ shows a continuous increase in the mean Eu valence, reaching +2.4 around 40 GPa \cite{Utsumi_ES_2021, dhami2023pressure}.

Here, we extend our investigation to EuIrGe$_3$, where the more extended Ir 5$d$ orbital and strong spin-orbit coupling may give rise to distinct pressure-induced electronic and structural responses. 

At ambient pressure, EuIrGe$_3$ exhibits antiferromagnetic ordering below $T_{\rm N}$ = 12.3 K, forming a longitudinal sinusoidally modulated magnetic structure.  Upon further cooling, a second magnetic transition occurs at $T_{\rm N}' = 7.5$ K, leading to a cycloidal magnetic structure, before the plane rotates by 45$^{\circ}$ below $T_{\rm N}^*$ = 5 K \cite{Matsumura2022}.
A previous high-pressure electrical resistivity measurement of EuIrGe$_3$ revealed a monotonic increase of magnetic ordering temperatures up to 8 GPa \cite{Kakihana2017}. However, direct observations of the evolution of Eu valence and the crystal structure under pressure have been lacking.
In the present work, we combine pressure-dependent XAS, synchrotron XRD complemented by density functional theory (DFT) calculations, and electrical resistivity measurements to investigate the evolution of the electronic and crystal structure of EuIrGe$_3$ under pressure. We demonstrate a continuous increase in the Eu valence with pressure, accompanied by pronounced modifications of the ligand electronic structure and anisotropic lattice compression. Furthermore, our high-pressure XRD measurements provide evidence for a structural phase transition above 38 GPa, establishing EuIrGe$_3$ as a distinct member of the BaNiSn$_3$-type Eu$T$Ge$_3$ family. In addition, electrical resistivity measurements reveal a linear increase in the magnetic transition temperatures up to 18 GPa, suggesting that the enhancement of magnetic interactions persists despite the continuous evolution toward an intermediate valence state.

\section{Experimental methods}
Single crystals of EuIrGe$_3$ were grown using the metal-flux technique with liquid indium as the solvent. Details of the crystal growth procedure and characterization are provided in Ref. \cite{Bednarchuk_JAC_2015,  Bednarchuk_JAC_2_2015, Bednarchuk_APPA_2015}. 

Near-edge XAS measurements in transmission mode were performed under pressures up to 54~GPa at the ODE beamline of Synchrotron SOLEIL. The beamline utilizes a dispersive XAS setup, in which the incident X-ray beam is focused onto the sample by a bent Si(111) crystal, producing a beam spot of approximately 30$\times$30~$\mu$m$^2$ (FWHM). The dispersive geometry enables rapid data acquisition by simultaneously recording a broad energy range. The transmitted X-rays were recorded using a position-sensitive charge-coupled device (CCD) detector \cite{baudelet2011ode}. 
The measurements were conducted on both single-crystalline samples (approximately 20~$\mu$m thick) and powder samples of EuIrGe$_3$ at room temperature and at 4~K using a helium-flow cryostat. 
The samples were loaded into a rhenium gasket with a sample chamber diameter of 150~$\mu$m in a gas-membrane-driven diamond anvil cell (DAC) equipped with 250~$\mu$m culet diamonds. A 4:1 methanol–ethanol mixture was used as the pressure-transmitting medium (PTM). 

Additional near-edge XAS measurements were performed in the high energy resolution fluorescence detected (HERFD) mode at the GALAXIES beamline of Synchrotron SOLEIL \cite{Rueff_2014, Ablett_2019}.
For the HERFD measurements, the sample was loaded into a symmetric screw-driven DAC using a beryllium gasket and neon as PTM. Pressure was applied by manually tightening the four screws on the DAC. 
For both transmission and HERFD XAS measurements, the pressure was determined from the ruby fluorescence before and after each XAS measurement at a given pressure.

Powder XRD measurements under pressure up to 43 GPa were performed at the PSICHE beamline of Synchrotron SOLEIL using an incident photon energy of 33 keV ($\lambda$ = 0.3738 \AA). The powder sample was loaded into a gas-membrane-driven DAC equipped with 300 $\mu$m culet diamonds. A rhenium gasket with a 150 $\mu$m diameter sample chamber was used, with gold powder serving as the pressure calibrant and neon as the PTM. The pressure was determined from the equation of state of gold \cite{heinz1984}. The pressure evolution of the crystal structure of EuIrGe$_3$ was also investigated theoretically using Quantum ESPRESSO \cite{Giannozzi2009,Giannozzi2017}. The details of computational methodology are described in Sec. \ref{sec:dos}.

Temperature-dependent electrical resistivity under pressure was measured on single crystals of EuIrGe$_3$ using a standard four-probe method. The sample was loaded into a stainless steel gasket electrically insulated with a mixture of $\alpha$-Al$_2$O$_3$ powder and two-component epoxy (Loctite Stycast 1266), with Daphne 7373 used as the PTM. A DAC equipped with 750~$\mu$m culet diamonds and a 350~$\mu$m sample chamber was used for the measurements. The pressure was determined from ruby fluorescence before and after each temperature cycle.

\begin{figure*}[!htb]
    \centering
    \includegraphics[width=\textwidth]{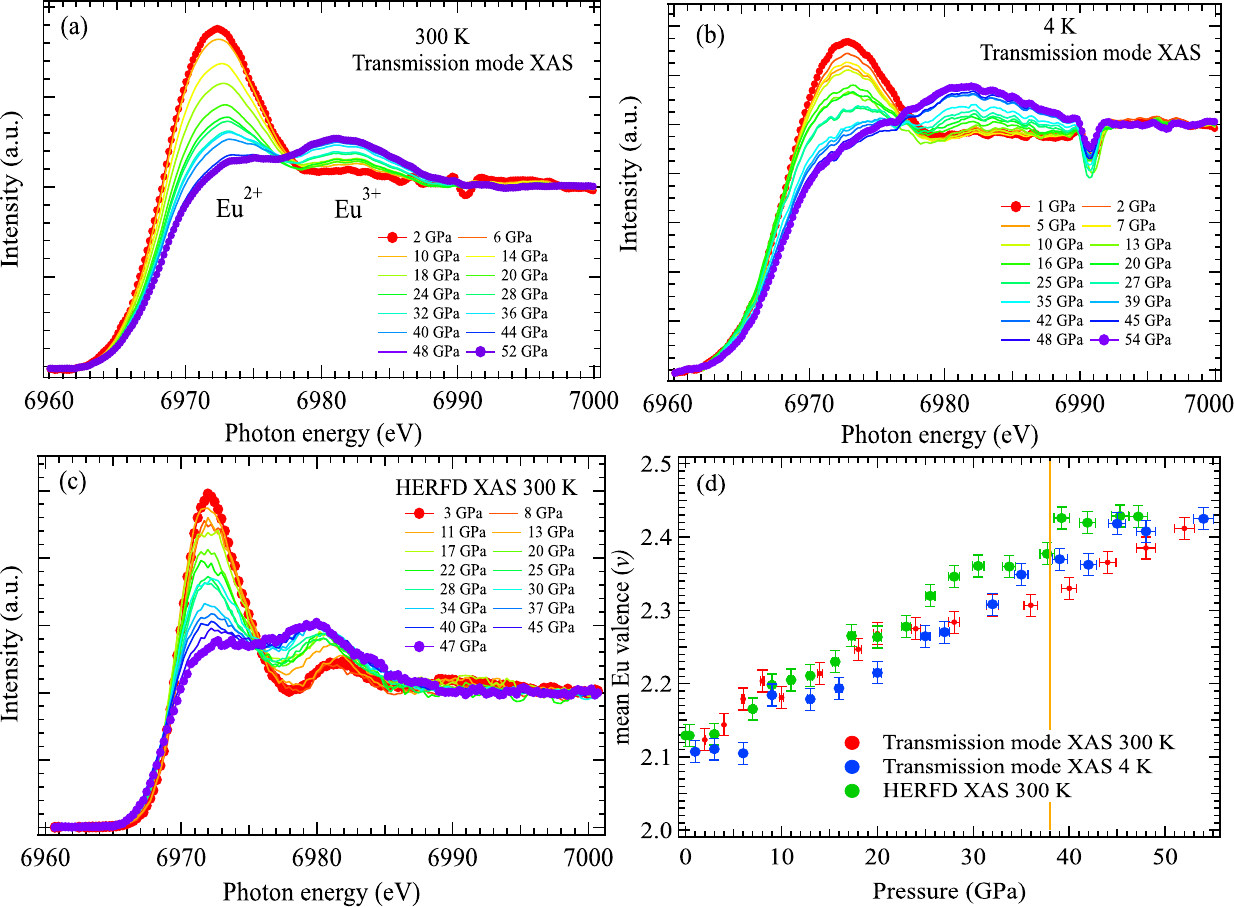}
    \caption{Pressure-dependent Eu $L_3$-edge XAS spectra of EuIrGe$_3$ measured in (a) transmission mode at 300 K, (b) transmission mode at 4 K, and (c) HERFD mode at 300 K. All spectra were normalized to the high energy region after subtraction of a constant pre-edge background. The feature at 6991 eV in the transmission spectra is an artifact of the beamline optics. (d) Pressure dependence of the mean Eu valence extracted from all measurements. The vertical orange line indicates the pressure at which the structural transition was observed (see Sec.\ref{subsec:xrd}).}
    \label{figEuXAS}
\end{figure*}

\section{Results and Discussion}

\subsection{Near-edge XAS under pressure}

Figure \ref{figEuXAS} shows pressure-dependent Eu $L_3$-edge XAS spectra of EuIrGe$_3$. In transmission geometry, incident ($I_0$) and transmitted ($I$) x-ray intensities were measured, and the absorption spectrum was obtained from their logarithmic ratio, ln($I_0$/$I$). The Eu $L_3$-edge XAS spectrum arises mainly from the dipole-allowed 2$p_{3/2}$ $\rightarrow$ 5$d$ transition. Although this transition does not directly probe the 4$f$ states, the strong Coulomb interaction between the 2$p$ core hole and the 4$f$ electrons gives rise to well-separated spectral features corresponding to the Eu$^{2+}$ (4$f^7$) and Eu$^{3+}$ (4$f^6$) configurations. At low pressures, the Eu $L_3$-edge XAS spectrum is dominated by the Eu$^{2+}$ peak at 6972 eV, with a small contribution from the Eu$^{3+}$ component centered at 6982 eV. Upon increasing pressure, the intensity of the Eu$^{2+}$  peak decreases, while that of the Eu$^{3+}$ peak increases.  The same behavior is observed in the Eu $L_3$-edge XAS spectra measured at 4 K, well below $T_{\rm N}$.

The pressure dependence of the Eu $L_3$-edge XAS spectrum was also investigated using the HERFD technique.  Like conventional XAS, Eu $L_3$-edge HERFD-XAS probes the dipole-allowed the 2$p_{3/2}$ $\rightarrow$ 5$d$ transition. However, the spectrum is recorded by scanning the incident photon energy across the Eu $L_3$ absorption edge while selectively detecting the Eu $L_{\alpha1}$ emission line ($2p^5 4f^n 5d^1 \rightarrow 2p^6 3d^9 4f^n 5d^1$). This detection scheme suppresses the 2$p$ core-hole lifetime broadening, resulting in substantially improved energy resolution. As shown in Fig. \ref{figEuXAS} (c), the low pressure HERFD-XAS spectrum is dominated by the Eu$^{2+}$ peak at 6972 eV, with a weak Eu$^{3+}$ peak centered at 6982 eV. These features are consistent with those observed in the transmission XAS spectra, but exhibit considerably sharper spectral profiles due to the improved energy resolution. Furthermore, the HERFD-XAS spectra show the same pressure evolution as the transmission XAS spectra, confirming the pressure-induced spectral change in EuIrGe$_3$.

The mean Eu valence ($\nu$) was extracted using a two-component fitting procedure following Refs.~\cite{Utsumi_ES_2021, dhami2023pressure}. At ambient pressure, the mean valence is $\nu = 2.1 \pm 0.02$, slightly deviating from the integer value. With increasing pressure, $\nu$ increases approximately linearly, reaching $2.40 \pm 0.02$ at 40~GPa, and then tends toward saturation above 40 GPa, as determined from HERFD-XAS measurements. The good agreement between the HERFD and transmission data obtained at 300 K supports the consistency of the valence analysis. Furthermore, no pronounced change in the Eu valence is observed between the transmission measurements at 300 and 4 K, suggesting that no temperature-driven valence transition occurs within the investigated pressure range.

\begin{figure}[!htb]
    \centering
\includegraphics[width=1.0\columnwidth]{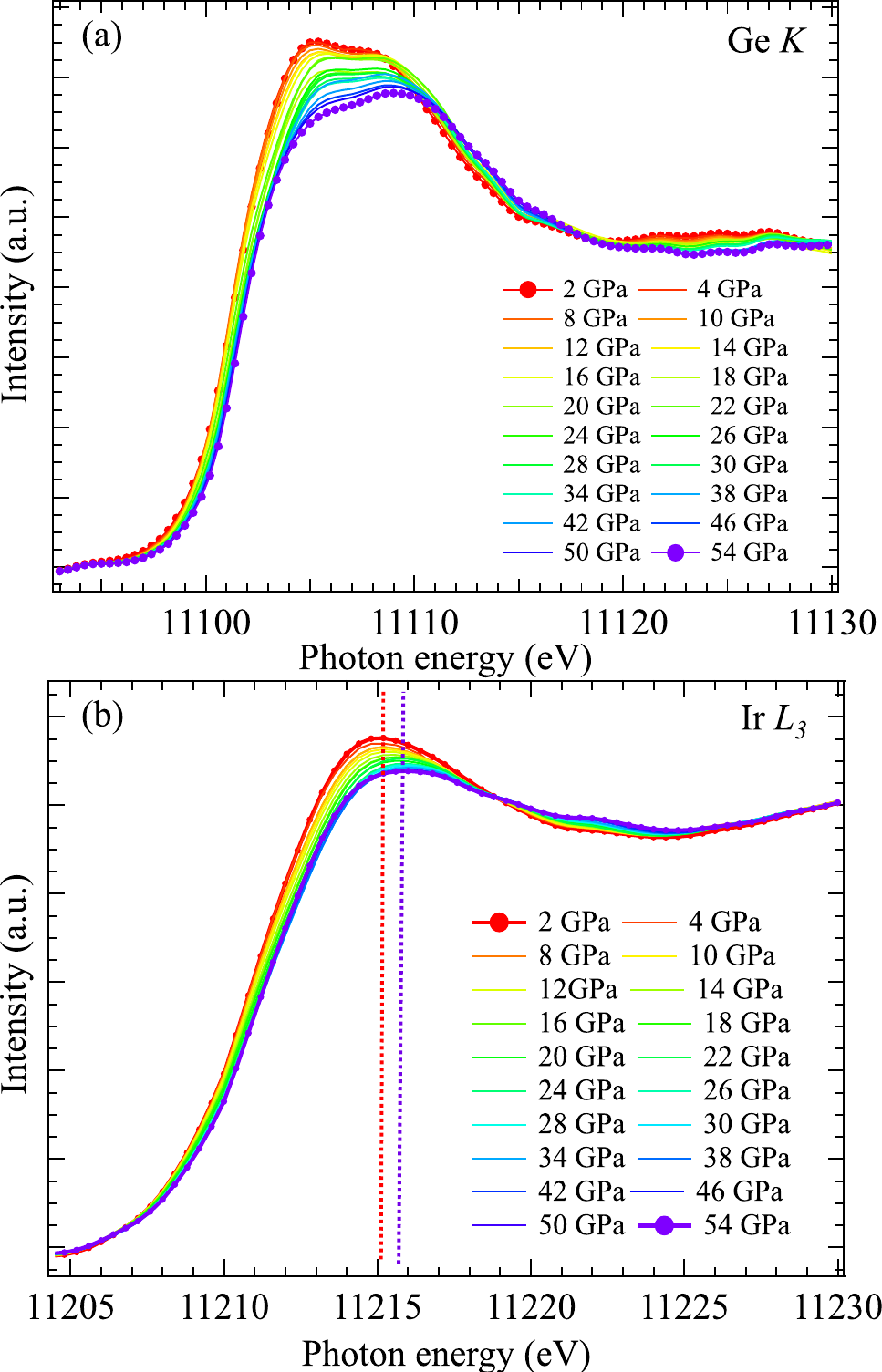}
\caption{Pressure-dependent (a) Ge $K$-edge and (b) Ir $L_3$-edge XAS spectra of EuIrGe$_3$ measured at 300 K in transmission mode. The spectra were normalized following the same procedure as for the Eu $L_3$-edge data. The red and violet vertical dashed lines indicate the peak positions at 2 and at 54 GPa, respectively, as determined from the Gaussian fits.}
    \label{figTrXASGe}
\end{figure}

To investigate the pressure-induced evolution of the electronic structure, Ge $K$-edge and Ir $L_3$-edge XAS measurements were also performed in transmission mode. Fig. \ref{figTrXASGe} (a) shows the Ge $K$-edge XAS spectra measured at 300 K. The spectrum exhibits two peaks at 11102~and 11106~eV. Double-peak features have also been observed in the Ge $K$-edge XAS spectra of the isostructural compounds CeCoGe$_3$~\cite{rogalev2021anisotropy} and EuCoGe$_3$ \cite{dhami2023pressure}, and have been attributed to Ge atoms occupying the $4b$ and $2a$ Wyckoff sites, respectively. In EuIrGe$_3$, DFT calculations show that the unoccupied states associated with both the Ge $4b$ and $2a$ sites exhibit two distinct components at 0 GPa (see Fig.\ref{figAP2}). With increasing pressure, these components broaden and shift toward higher energies, accompanied by a gradual redistribution of the calculated density of states. Experimentally, the overall XAS spectral intensity decreases and the spectral features broaden with increasing pressure. Concurrently, the higher-energy peak gradually gains intensity relative to the low-energy peak. Such pronounced pressure-induced spectral changes were not observed in EuCoGe$_3$, which exhibits only a small change in the Eu valence with increasing pressure \cite{dhami2023pressure}. The distinct evolution of the Ge $K$-edge spectra in EuIrGe$_3$ therefore suggests a stronger pressure-induced modification of the Ge-derived unoccupied electronic states, consistent with the more pronounced pressure-induced changes in the Eu valence. 

Figure \ref{figTrXASGe} (b) shows the pressure-dependent Ir $L_3$-edge ($2p_{3/2} \rightarrow 5d$) XAS spectra measured at 300 K. The spectra exhibit a pronounced white-line peak at approximately 11215 eV, close to the reported peak position of metallic Ir \cite{Clancy2012}. With increasing pressure, the peak shifts systematically toward higher energy up to $\sim$40 GPa, above which the shift tends to saturate. In general, the Ir $L_3$-edge peak energy is sensitive to the occupancy of the Ir 5$d$ states, with a higher peak energy typically associated with an increase in the Ir oxidation state ($i.e.$ larger number of 5$d$ holes). Clancy $et~ al$. \cite{Clancy2012} reported $\sim$1.3 eV shift of the Ir $L_3$-edge white-line peak per additional 5$d$ hole across a series of Ir compounds. The observed energy shift in EuIrGe$_3$ is smaller than 1 eV and is accompanied by a gradual decrease in the peak intensity. These changes suggest that pressure primarily modifies the Ir 5$d$ electronic structure through modifications of the local bonding environment, rather than inducing a change in the Ir oxidation state.
In addition to the main peak, subtle changes are observed in the shoulder feature near 11224~eV. This feature is likely associated with multiple scattering contributions to the near-edge XAS spectrum, and its pressure evolution suggests modifications of the local Ir-Ge bonding environment.

The enhanced pressure-induced Eu valence evolution in EuIrGe$_3$ compared with EuCoGe$_3$ \cite{dhami2023pressure},  and its similarity to EuRhGe$_3$\cite{Utsumi_ES_2021}, can be understood in terms of the increasing spatial extent and hybridization strength of the transition-metal $d$ orbitals across the 3$d$, 4$d$, and 5$d$ series. The more extended 5$d$ orbitals of Ir provide stronger overlap with Ge $4p$ states, leading to a more strongly hybridized Ir-Ge electronic framework. This enhanced hybridization can promote stronger coupling between conduction bands involving Ir 5$d$, Ge 4$p$, and Eu $5d$ orbitals  with the localized Eu $4f$ states. The stronger Eu valence response observed in EuIrGe$_3$ compared with the Co and Rh analogues therefore likely reflects the enhanced Ir--Ge hybridization and its role in promoting stronger Eu 4$f$--conduction band hybridization under pressure.

\subsection{XRD measurements under pressure}
\label{subsec:xrd}
\begin{figure*}[!htb]
    \centering
    \includegraphics[width=\textwidth]{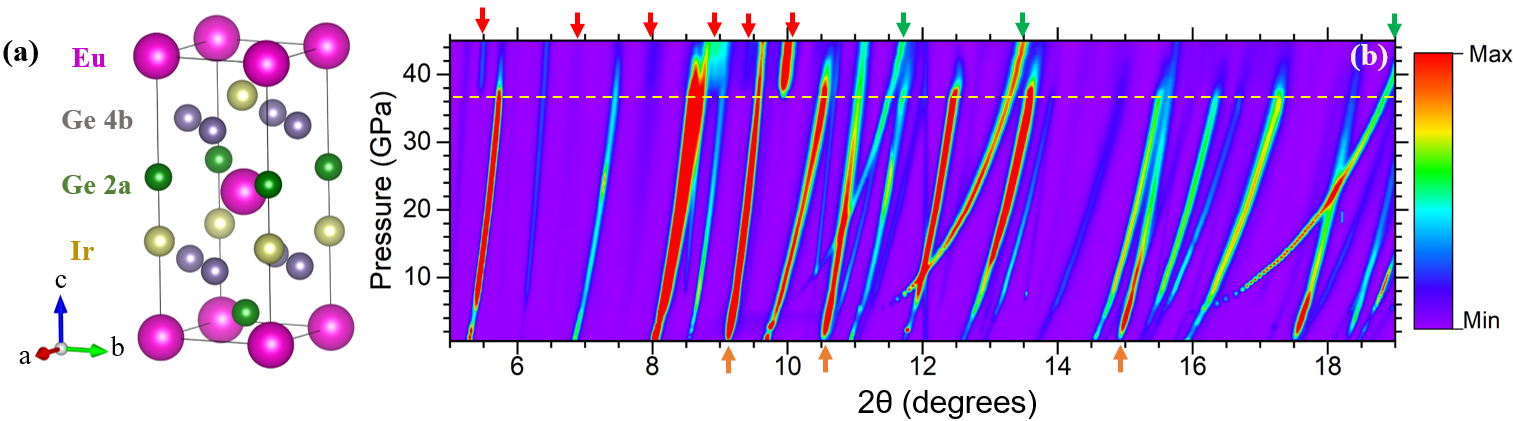}
    \caption{(a) Crystal structure of EuIrGe$_3$. The Ge 2$a$ and Ge 4$b$ sites are shown in different colors for clarity. The structure was drawn using VESTA \cite{momma2008vesta}. (b) Contour map of the powder XRD intensity of EuIrGe$_3$ measured at room temperature as a function of pressure. The rapidly shifting peaks, indicated by green arrows, originate from the solidification of the neon pressure transmitting medium \cite{Finger1981}. The emergence of additional diffraction peaks above $\sim$38 GPa indicates a possible structural phase transition. Diffraction peaks from the gold pressure calibrant are marked by golden arrows, while those associated with the high-pressure phase are indicated by red arrows.}
    \label{fig3}
\end{figure*}

Powder XRD measurements were performed on EuIrGe$_3$ at room temperature under pressure up to 43 GPa. The contour map of the diffraction intensity over the pressure range from 1 to 43 GPa is presented in Fig. {\ref{fig3}}(b). In order to highlight the pressure evolution of the Bragg peaks, the contour map is plotted in the 2$\theta$ range from 5 to 19$^\circ$. The diffraction pattern contains reflections from the main phase EuIrGe$_3$, gold used as pressure calibrant, and the neon PTM. Rietveld refinements were performed for all three phases using Profex \cite{Profex}. Representative diffraction patterns and corresponding Rietveld refinement profiles at selected pressures are presented in supplementary material.

At ambient pressure and 300 K, EuIrGe$_3$ crystallizes in the tetragonal $I4mm$ (107) space group with lattice parameters \textit{a} = 4.4418 Å, \textit{c} = 10.068 Å, and a unit-cell volume of $V$ = 198.65 Å$^3$. High-pressure XRD measurements show that the tetragonal $I4mm$ structure is retained up to at least 38 GPa. Upon further compression, the progressive weakening of the existing diffraction peaks, accompanied by the emergence of additional reflections, suggests the onset of a structural phase transition. The saturation of the mean Eu valence above $\sim40$~GPa coincides with the onset of the structural transition, suggesting a strong coupling between the Eu valence and the lattice under compression. The increase in Eu$^{3+}$ character with decreasing volume is consistent with the smaller ionic size of Eu$^{3+}$ relative to Eu$^{2+}$ \cite{shannon1976revised}. The structural transition may therefore modify the local Eu coordination and Eu-ligand hybridization, thereby limiting the further change of the Eu valence at higher pressures. This behavior highlights the close interplay between the electronic structure and lattice degrees of freedom under compression.

The coexistence of diffraction peaks from both phases is consistent with a transition to a lower-symmetry orthorhombic or monoclinic structure. However, further high-pressure single-crystal XRD measurements above 38 GPa are required to determine the crystal structure of the high-pressure phase and unambiguously establish the nature of the structural transition. 

\begin{figure*}[!htb]
    \centering
    \includegraphics[width=\textwidth]{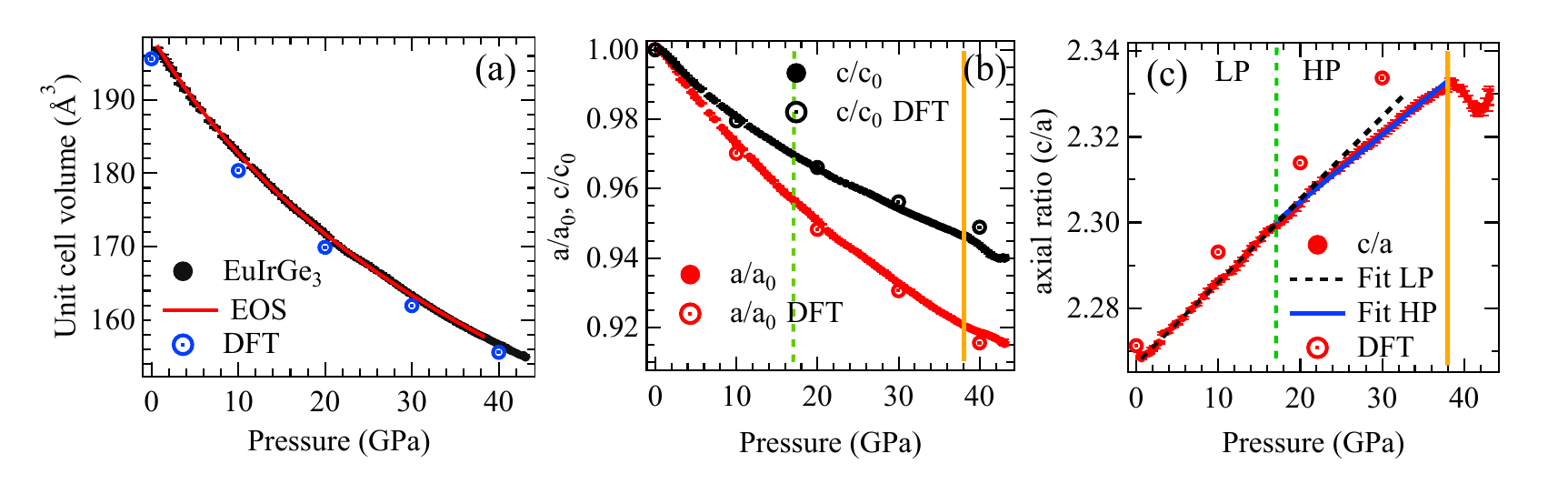}
    \caption{(a) Pressure dependence of the unit-cell volume of EuIrGe$_3$ (black circles), together with the EOS fit (solid line). Unit-cell volumes obtained from DFT calculations at selected pressures (open blue circles) are also shown. Error bars are smaller than the symbol size. (b) Relative changes in the lattice parameters of EuIrGe$_3$, normalized to their values at the lowest experimental pressure, as a function of pressure. The orange vertical bar indicates the pressure at which the structural phase transition occurs. DFT results are shown as open circles. (c) Pressure dependence of the axial ratio, $c/a$. Linear fits to the LP ($\leq 17$ GPa; dashed black line) and HP (17--38 GPa; solid blue line) regions are shown. To highlight the change in slope near 17 GPa, the LP linear fit is extrapolated into the HP region. The vertical green dashed line marks the crossover pressure between the LP and HP regimes. The DFT results deviate from the experimental trend at higher pressures.
}
    \label{fig4}
\end{figure*}

Figure \ref{fig4} (a) shows the pressure dependence of the unit cell volume of EuIrGe$_3$ obtained from powder XRD measurements. The unit cell volumes calculated by DFT at selected pressures are also shown and are in good agreement with the experimental results up to the highest pressure investigated. In order to study the elastic properties of EuIrGe$_3$, the pressure dependence of the unit cell volumes was fitted using the EOSFit7c software \cite{EosFit}. The fitting was performed using the 3$^{rd}$ order Birch Murnaghan equation of state (EOS) \cite{Birch1947} up to 38 GPa. The resulting EOS parameters are summarized in Table \ref{table2} together with those reported for the related Eu$T$Ge$_3$ compounds. The bulk modulus obtained for EuIrGe$_3$, $B_0 = 100.01 (2)$~GPa, is appreciably larger than those reported for the isostructural EuCoGe$_3$ and EuRhGe$_3$ compounds, $B_0 = 75.6$ \cite{dhami2023pressure} and $73(1)$~GPa \cite{dhami2024synchrotron}, respectively. The enhanced incompressibility is also reproduced by DFT calculations, which yield $B_0 = 105.9(1)$~GPa for EuIrGe$_3$. This behavior may be related to the stronger hybridization involving spatially extended Ir-$5d$ and Ge 4$p$ states, resulting in a stiffer Ir--Ge bonding network.

\begin{table}[htb!]
\centering
\begin{tabular}{c c c c}
 \hline
Compound & (V$_0$) & ($B_{0}$) & $B_{0}^\prime$\\

 \hline
EuCoGe$_3$ & 184.39 (Å$^3$)& 75.6 & 5.58 \cite{dhami2023pressure}\\ 
EuNiGe$_3$  & 185.7 (Å$^3$)& 79 & 8.8 \cite{chen2023evidence}\\ 
EuRhGe$_3$ & 196.50 (Å$^3$)& 73 (1) & 5.5 (2) \cite{dhami2024synchrotron}\\
EuIrGe$_3$ (XRD) & 198.65 (Å$^3$)& 100.01 (2) & 4.34 (1) \\
EuIrGe$_3$ (DFT) & 195.61 (Å$^3$)& 105.9 (1) & 4.4  \\
\hline
\end{tabular}
\caption{The unit cell volume ($V_0$), bulk modulus ($B_0$), and first pressure derivative of bulk modulus $B_{0}^\prime$ of Eu$T$Ge$_3$ series and those obtained by the DFT calculation in EuIrGe$_3$. The values of EuCoGe$_3$, EuNiGe$_3$ and EuRhGe$_3$ are taken from Ref. \cite{dhami2023pressure}, \cite{chen2023evidence} and \cite{dhami2024synchrotron}, respectively.}
\label{table2}
\end{table}

\begin{figure*}[!htb]
    \centering
    \includegraphics[width=1.0\textwidth]{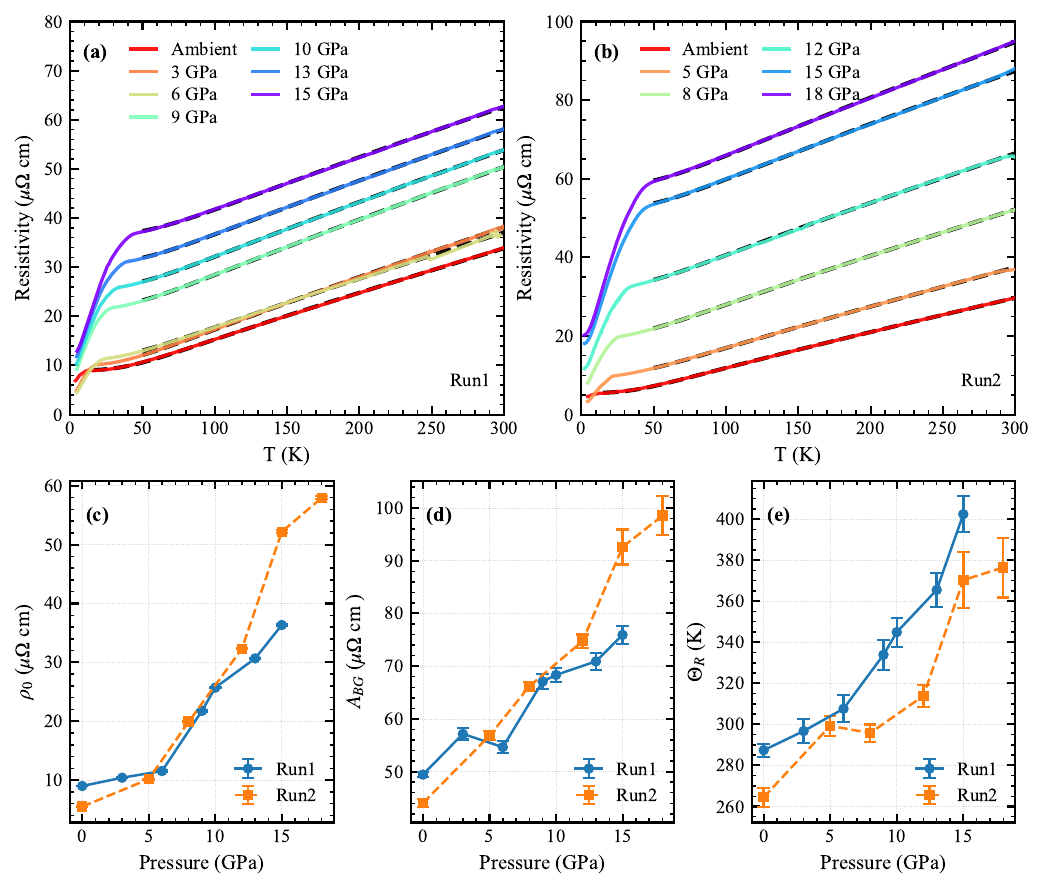}
    \caption{Temperature-dependent electrical resistivity of EuIrGe$_3$ measured at various pressures during (a) Run1 and (b) Run2. The two runs were performed on different crystals from the same growth. The dashed black lines represent fits to the BG model over the temperature range 50--300 K at high pressures and 15--300 K at ambient pressure. Pressure dependence of the fitted (c) residual resistivity, $\rho_0$, (d) BG coefficient, $A_{\mathrm{BG}}$, and (e) BG temperature, $\Theta_R$. }
    \label{figRho}
\end{figure*}

Figure \ref{fig4} (b) shows the relative changes in the lattice parameters of EuIrGe$_3$ as a function of pressure. The lattice parameters $a$ and $c$ are normalized to their values at the lowest experimental pressure. The $a$ axis exhibits greater compressibility than the $c$ axis. Similar pressure-dependent anisotropic compression has also been observed in EuCoGe$_3$ \cite{dhami2023pressure}, EuRhGe$_3$ \cite{dhami2024synchrotron} and EuNiGe$_3$ \cite{chen2023evidence},  suggesting that this behavior is an intrinsic character of the Eu$T$Ge$_3$ family. Around 38 GPa, the $c$-axis exhibits a pronounced change in its compressibility, suggesting the onset of a structural phase transition.

Fig. \ref{fig4} (c) shows the pressure dependence of the axial ratio ($c/a$) of EuIrGe$_3$. The $c/a$ ratio exhibits a change in slope, which can be divided into two regions: a low-pressure (LP) region \textless17 GPa and a high-pressure (HP) region above 17 GPa. Linear fits to the LP and HP regions yield $c/a$= 2.267+0.002\textit{P} and $c/a$= 2.273+0.0015\textit{P}, respectively. Similar changes in the pressure dependence of $c/a$ have also been observed in EuCoGe$_3$ \cite{dhami2023pressure} and EuRhGe$_3$ \cite{dhami2024synchrotron}, where the magnitude of the slope change is even larger, despite no structural transition being reported. This comparison suggests that the kink in $c/a$  at 17 GPa is not directly associated with the structural transition. Instead, it likely reflects pressure-induced modifications of the electronic structure and chemical bonding within the tetragonal phase.  At approximately 38 GPa, the $c/a$ ratio also shows a clear deviation from its preceding pressure dependence, consistent with the structural phase transition observed at this pressure.
The structural transition in EuIrGe$_3$ may be associated with pressure-induced electronic or lattice instabilities related to the distinct electronic structure of the Ir-based compound, including stronger spin-orbit coupling and more spatially extended Ir 5$d$ orbitals.

{\subsection{Electrical resistivity under pressure}

The temperature dependence of the electrical resistivity of EuIrGe$_3$ exhibits metallic behavior. As shown in Fig. \ref{figRho}, the resistivity decreases sharply below the antiferromagnetic transition temperatures, $T_{\rm N}\sim$12 K and  $T_{\rm N}'\sim$7.5 K, owing to the suppression of spin-disorder scattering associated with the ordering of the Eu moments. Although an additional transition near 5 K has previously been reported from magnetization measurements \cite{Maurya2016, Kakihana2017}, no corresponding anomaly is resolved in the resistivity data presented here. 

\begin{figure*}[!htb]
    \centering
    \includegraphics[width=1.0\textwidth]{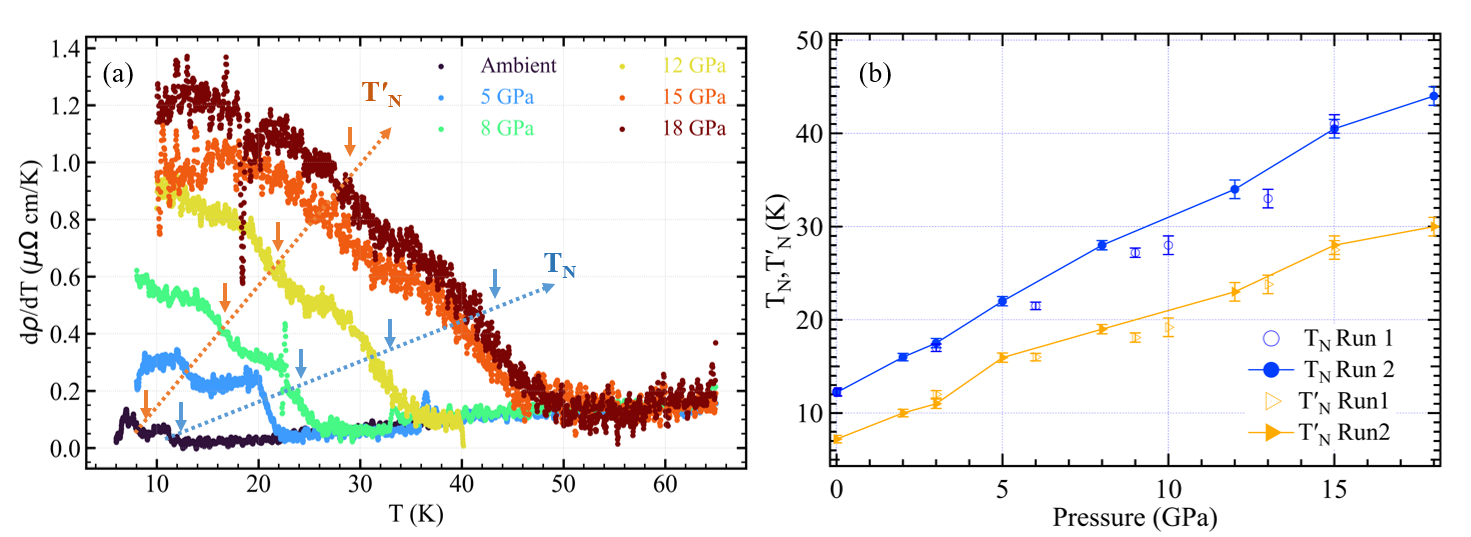}
    \caption{\textbf{(a)} Temperature derivative of the electrical resistivity, $d\rho/dT$, of EuIrGe$_3$ measured under pressure during Run2. The derivative curves were smoothed to facilitate the identification of the magnetic transition temperatures. \textbf{(b)} Pressure dependence of the magnetic ordering temperatures of EuIrGe$_3$. Open (filled) circles denote $T_{\mathrm{N}}$ obtained from Run1 (Run2), while open (filled) triangles represent $T_{\mathrm{N}}'$ from Run1 (Run2).}
    \label{figR1}
\end{figure*}

The electrical resistivity of EuIrGe$_3$ was measured under pressure up to 18 GPa. The magnetic transition temperatures were determined from the first temperature derivative of the resistivity and are shown in Fig.~\ref{figR1}. The resistivity data indicates that both magnetic ordering temperatures, $T_{\rm N}$ and $T_{\rm N}'$, increase monotonically with pressure. This behavior is consistent with the earlier study by Kakihana $et~al.$ \cite{Kakihana2017}, while extending the investigated pressure range from 8 to 18 GPa. Since electrical resistivity does not directly probe the magnetic structure, possible pressure-induced changes in the magnetic ordering cannot be ruled out. It is worth mentioning that the isostructural compound EuIrSi$_3$ orders antiferromagnetically at $T_{\rm N}$ = 52 K \cite{maurya2022large} at ambient pressure. Because Si is isovalent with Ge but has a smaller atomic radius, substituting Si for Ge exerts positive chemical pressure, leading to a reduced unit cell volume. Indeed, the ambient pressure unit cell volume of EuIrSi$_3$ is comparable to that of EuIrGe$_3$ under high pressure \cite{maurya2022large}. This correspondence is consistent with the observed enhancement of the magnetic ordering temperatures in EuIrGe$_3$ under applied pressure. 

In addition to the increase in the magnetic ordering temperatures, the electrical resistivity increases with pressure over the entire measured temperature range. This behavior contrasts with the trend generally expected for a simple metal, in which pressure increases the electronic bandwidth and thereby tends to reduce the electrical resistivity \cite{goree1966pressure, rapp1981pressure}. A pressure-induced increase in electrical resistivity has nevertheless been reported in several Eu-based intermetallic compounds, including EuRh$_2$Si$_2$ \cite{honda2018pressure}, EuRhSi$_3$ \cite{Nakashima_JPSJ_2017}, and EuCu$_2$Ge$_2$ \cite{gouchi2020quantum}. The microscopic origins of this anomalous pressure dependence may differ among these systems and have been discussed in terms of pressure-induced changes in electronic structure, Eu valence, and electronic correlations. To determine whether the resistivity increase in EuIrGe$_3$ can be attributed to changes in the phonon contribution, we first analyze the high-temperature resistivity (50--300 K) using the Bloch--Grüneisen formalism \cite{bloch1930,gruneisen1933} (see equation \ref{eq1}).
\begin{equation}
\rho(T)=\rho_0 + A_{\rm{BG}}\left(\frac{T}{\Theta_R}\right)^5
\int_{0}^{\Theta_R/T}
\frac{x^5 e^{x}}{\left(e^{x}-1\right)^2}\,dx
\label{eq1}
\end{equation}

The fits reveal a pronounced increase in the resistive Bloch--Grüneisen
temperature, $\Theta_R$, from approximately $270$~K at ambient pressure
to approximately $370$~K at $18$~GPa, indicating an increase in the effective phonon energy scale relevant to electron--phonon scattering (see Fig. \ref{figRho}). The increase in $\Theta_R$ is
accompanied by a substantial increase in the fitted temperature-independent contribution, $\rho_0$, particularly at higher pressures. Within the Bloch--Grüneisen description, an increase in $\Theta_R$
would, for otherwise comparable $A_{\rm BG}$, reduce the phonon
contribution to the resistivity at a given temperature. Thus, the
pronounced increase in the measured resistivity with pressure cannot
be accounted for solely by pressure-induced lattice stiffening and instead points to an additional pressure-dependent contribution to the scattering.

The gradual increase in the mean Eu valence, together with the pressure-induced evolution of the Ge $K$-edge and Ir $L_3$-edge XAS spectra, indicates substantial changes in the electronic structure under compression. These changes may contribute to the additional pressure-dependent scattering and hence to the anomalous increase in resistivity observed over a broad temperature range.

\section{Conclusion}
The pressure-induced evolution of the electronic and crystal structures of EuIrGe$_3$ was investigated by combining XAS, synchrotron XRD, DFT calculations and electrical resistivity measurements.  The Eu $L_3$-edge XAS reveals continuous increase in the mean Eu valence from +2.1 at ambient to 2.4+ around 40 GPa, with a tendency towards saturation at higher pressures. Concurrent shifts and spectral-shape changes at the Ir $L_3$ and Ge $K$ edges indicate substantial  pressure-induced modifications of the Ir and Ge electronic  states and associated charge redistribution. 
  
High-pressure XRD reveals anisotropic lattice compression and a structural phase transition around 38 GPa. The relatively large bulk modulus of EuIrGe$_3$ may be related to a stronger hybridization involving the spatially extended Ir 5$d$ and Ge states.

Electrical resistivity measurements show a monotonic increase in the magnetic transition temperatures up to 18 GPa, indicating a robust antiferromagnetic ground state under compression. At the same time, the increase in resistivity cannot be accounted for solely by lattice stiffening, suggesting an additional contribution from pressure-induced changes in the electronic structure and scattering mechanisms.

Overall, these results demonstrate a strong interplay among Eu valence, electronic structure, lattice degrees of freedom, and magnetism in EuIrGe$_3$, and highlight the important role of the transition-metal $d$ states in the pressure response of noncentrosymmetric Eu$T$Ge$_3$ compounds.

\section*{Acknowledgments}
N.S.D. gratefully acknowledges valuable and fruitful discussions with Neven Barišić. The high pressure experiments at Synchrotron SOLEIL were performed under proposal Nos. 20231234 (ODE), 20210673 (GALAXIES), 99220002 (PSICHE), and 20220242 (CRISTAL).
This work was supported by the Croatian Science Foundation under project No. UIP-2019-04-2154 and, in part, by project Nos. IP-2020-02-9666 and IP-2025-02-7384. The work at the Institute of Physics was further supported by the project Cryogenic Centre at the Institute of Physics -- KaCIF co-financed by the Croatian Government and the European Union through the European Regional Development Fund-Competitiveness and Cohesion Operational Programme (Grant No. KK.01.1.1.02.0012), and the project Ground states in competition – Strong Correlations, Frustration and Disorder — FrustKor, financed by the Croatian Government and the European Union through the National Recovery and Resilience Plan 2021-2026 (NPOO). V.B. acknowledges the Paris Ile-de-France Region in the framework of DIM MaTerRE (project DAC-VX).

\bibliographystyle{apsrev}
\bibliography{Biblography_EuIrGe3}

\clearpage

\appendix
\onecolumngrid

\section{DFT calculated DOS under pressure}	
\label{sec:dos}

Density functional theory (DFT) calculations were performed using the plane-wave pseudopotential method as implemented in the Quantum ESPRESSO package (v7.2) \cite{Giannozzi2009, Giannozzi2017}, employing the Perdew-Burke-Ernzerhof functional revised for solids (PBEsol) \cite{Perdew2008} for the exchange-correlation potential and optimized pseudopotentials from the pslibrary-1.0.0 database \cite{DALCORSO2014337}. Strong electron correlation effects on the localized Eu 4$f$  were treated within the DFT+U framework \cite{Marzari1999, timrov2022hp}. The Hubbard interaction $U$ for Eu was assumed to be 3.8 eV in order to match the Eu 4$f$ peak position observed in the photoelectron spectroscopy \cite{Utsumi2018}. The $U$ value was fixed to a constant throughout the whole pressure region. The Fermi-surface smearing was handled using the Marzari-Vanderbilt cold-smearing scheme \cite{Marzari1999}, with a smearing width of 0.003 Ry at all pressures. The wave functions and charge density/potential were expanded using kinetic energy cutoffs of 150 Ry and 700 Ry, respectively, and the Brillouin zone was sampled with 12×12×6 and 20×20×9 Monkhorst-Pack k-point meshes. Spin-orbit coupling was included for Eu. 

Figure \ref{figAP2} presents the calculated density of states (DOS) of EuIrGe$_3$. The Eu 4$f$ states are localized around -1 eV, while the unoccupied 
Eu states are predominantly of 5$d$ character. The Ir 5$d$ states extend from the Fermi level to approximately -6 eV. The occupied Ge 4$s$ states locate from -7 to -12 eV and are partially hybridized with the Ir 5$d$ states (not shown here). The Ge 4$p$ states are broadly distributed across the valence and conduction bands, with substantial overlap with both the Ir 5$d$ and Eu 5$d$ states.

\begin{figure*}[!htb]
    \centering
    \includegraphics[width=1.0\textwidth]{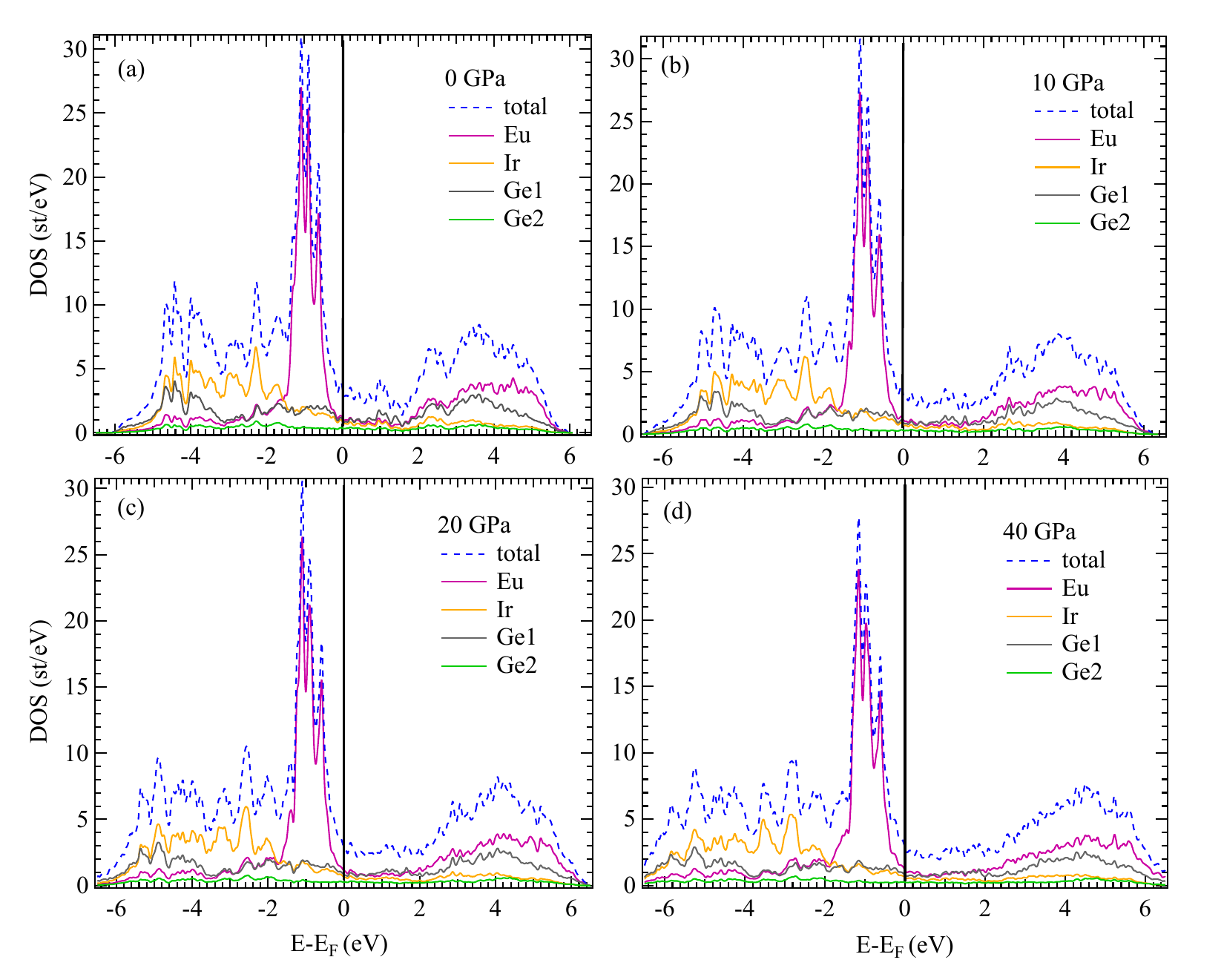}
    \caption{DFT calculated DOS of EuIrGe$_3$ at (a) 0 GPa, (b) 10 GPa, (c) 20 GPa, and (d) 40 GPa. Ge1 and Ge2 correspond to the Ge atoms occupying the 4$b$ and 2$a$  Wyckoff sites, respectively.}
    \label{figAP2}
\end{figure*}

\end{document}